\documentclass{article}
\usepackage{emulateapj,psfig}

\begin{document}

\title{``Late prompt'' emission in Gamma Ray Bursts? }

\normalsize \author{G. Ghisellini, G. Ghirlanda, L. Nava\altaffilmark{1}, 
C. Firmani\altaffilmark{2}} 
\affil{
INAF --
Osservatorio Astronomico di Brera, via Bianchi 46, I--23807 Merate, Italy}
\altaffiltext{1}{Univ. dell'Insubria, V. Valleggio, 11, I--22100, Como, Italy}
\altaffiltext{2}{Instituto de Astronom\'{\i}a, U.N.A.M., A.P. 70-264, 
04510, M\'exico, D.F., M\'exico}

\begin{abstract}
The flat decay phase in the first 10$^2$--10$^4$ seconds of
the X--ray light curve of Gamma Ray Bursts (GRBs) has not yet found
a convincing explanation.
The fact that the optical and X--ray lightcurves
are often different, with breaks at different times,
makes contrived any explanation based on the same origin
for both the X--ray and optical fluxes.
We here assume that the central engine can 
be active for a long time, producing shells 
of decreasing bulk Lorentz factors $\Gamma$.
We also assume that the internal dissipation of these late
shells produces a continuous and smooth emission (power--law in time),
usually dominant in X--rays and sometimes in the optical.
When $\Gamma$ of the late shells is larger than $1/\theta_j$, 
where $\theta_j$ is the jet opening angle, we see only a portion
of the emitting surface. 
Eventually, $\Gamma$ becomes smaller than $1/\theta_j$,
and the entire emitting surface is visible.
Thus there is a break in the light curve when $\Gamma=1/\theta_j$,
which we associate to the time at which the plateau ends.
After the steeply decaying phase which follows the 
early prompt, we see the sum of two emission components:
the ``late--prompt'' emission (due to late internal dissipation), 
and the ``real afterglow'' emission (due to external shocks).
A variety of different optical and X--ray light curves are then
possible, explaining why the X--ray and the optical light curves
often do not track each other (but sometimes do), 
and often they do not have simultaneous breaks.
\end{abstract} 


\section{Introduction}

One of the puzzling results of the Swift satellite
(Gehrels et al., 2004) is the discovery that the [0.3--10 keV] 
X--ray light curve of Gamma Ray Bursts (GRBs) is much more 
complex than thought in the pre--Swift era.
After a steep decline of the flux [$F(t) \propto t^{-\alpha_1}$,
with $\alpha_1 \sim$ 3--5; Tagliaferri et al. 2005], 
which is most commonly interpreted as off axis radiation 
of a switching--off fireball (see e.g. Kumar \& Panaitescu 2000), 
the flux decay becomes shallow [$F(t) \propto t^{-\alpha_2}$, 
with $\alpha_2\sim$ 0.2--0.8], 
up to a break time of $10^3$--10$^4$ s (Willingale et al. 2007,
hereafter W07), 
after which the flux decays ``normally'' [$F(t) \propto t^{-\alpha_3}$
with $\alpha_3\sim$ 1--1.5; Nousek et al. 2005], 
i.e. in a similar way as observed in the pre--Swift era.
In addition, several bursts show flares superposed to this power law
evolution (Burrows et al. 2005), 
leading Fan \& Wei (2005) and Lazzati \& Perna (2006) 
to suggest a long lasting central engine.
Unpredicted beforehand, the complex structure of the X--ray light
curve, characterized by a steep--flat--steep behavior, 
has been interpreted in several ways
(for reviews, see e.g. Panaitescu 2007; Granot 2007; Zhang 2007)
none of which seems conclusive.
The three main possibilities already proposed (but there are more, see
the review by Zhang 2007), are:
i) energization of the forward shock by the arrival of shells being 
produced late (with large $\Gamma$s), or just after the prompt phase 
(with small $\Gamma$s);
ii) changing microphysical parameters, assuming that the efficiency
of the forward shock to produce radiation increases with time:
iii) off--axis jets, whose prompt and early afterglow radiation
is not fully beamed towards the observer.
Note that the spectral slope does not change across the temporal
break from the shallow decay phase to the normal decay phase,
ruling out the crossing of a spectral break across the band.
This favors instead an hydrodynamical or geometrical nature of the break.
All these ideas do not obviously distinguish between X--ray and optical
radiation, which should have the same origin.
As a consequence, the light curves in both bands should be similar,
contrary to what observed in several cases
(see e.g. Panaitescu et al. 2006; Panaitescu 2007).

These difficulties recently led Uhm \& Beloborodov (2007)
and Grenet, Daigne \& Mochkovitch (2007) to consider the
possibility that the X--ray plateau emission is not due to the
forward, but to the reverse shock running into ejecta of
relatively small Lorentz factors.
This however requires an appropriate $\Gamma$--distribution of the
ejecta, and also the suppression of the X--ray flux produced by the 
forward shock.

Here we make the alternative proposal that the plateau phase
of the X--ray emission (and sometimes even of the optical) is due
to a ``late--prompt'' mechanism: after the early prompt (the prompt
which we are used to) there may be a tail of activity of the 
central engine, producing for a long time (i.e. days)
shells of progressively lower power and bulk Lorentz factor.
The dissipation process during this and the early phases can 
occur at similar radii.
The total energetics involved in this late activity phase is
smaller than (and at most comparable to) the energetics of the early
phase, but diluted on a much longer time.
The reason for the shallow decay phase, and for the break ending it,
is that the $\Gamma$--factor is decreasing, allowing to see an
increasing portion of the emitting surface, until all of it is
visible.

\section{Late prompt emission?}

Assume that the central engine, after having emitted 
most of the power in the usual duration
of what we call ``prompt'' emission,
continues to create shells of much smaller power, but 
for a much longer time.
For simplicity, let us call ``early prompt'' and ``late prompt''
the two phases of activity.
By contrast, we call ``real afterglow'' the emission produced
in the forward shock created by the interaction of the
shells with the circumburst medium.

The early prompt emission is due to internal dissipation
of shells of large $\Gamma$--factors 
(changing erratically)
and energy,
due to e.g. internal shocks (Rees \& Meszaros, 1994) or interactions
with the funnel of the progenitor star (Thompson 2006;
Thompson Meszaros \& Rees 2006), or some form of magnetic reconnection
(e.g. Spruit, Daigne \& Drenkhahn 2001).

We suggest that the late prompt emission
can 
be due to the same dissipation processes, but by shells created 
at late times with smaller $\Gamma$ and much lower power.
The radiation can then be produced at distances relatively
close to the central engine (even less than $10^{13}$--$10^{14}$ cm),
in a different region where the shells, produced during the early prompt,
interacts with the circumburst medium producing the real afterglow.
Note that a smaller $\Gamma$ implies less Doppler time contraction,
and therefore a less pronounced variability during the late prompt.
Furthermore,
if $\Gamma$ is decreasing with time, a new effect appears.
In fact, when $\Gamma >1/\theta_j$,
the emission surface seen by the observer is of the order
of $(R/\Gamma)^2$ (here $R$ is the distance from the black hole
where dissipation takes place, and $\theta_j$ is the jet opening angle),
and becomes $(\theta_j R)^2$ when $\Gamma \le 1/\theta_j$.
In the same way as in the afterglow case, we should then see a steepening
of the light curve when the central engine produces shells with 
$\Gamma\sim 1/\theta_j$.
This should occur at the time $t_a$, in a similar way as 
the jet break time $t_j$ for the afterglow.
According to this scenario, there is a link
between $t_a$ and $t_j$: both are times at which $\Gamma=1/\theta_j$,
but they refer to two different processes.

The plateau phase of the X--ray emission is characterized
by a power law decay $L(t)\propto t^{-\alpha_2}$,
followed by a steeper decay $L(t)\propto t^{-\alpha_3}$.
The end of the plateau phase occurs at $t_a$, and the transition
is smooth.
If $j^\prime$ is the bolometric emissivity in the comoving frame, we have
\begin{eqnarray}
L( t< t_a) &\sim&  \pi \left( {R\over \Gamma}\right)^2  
\Delta R^\prime \Gamma^2 j^\prime(t) \propto t^{-\alpha_2}, 
\,\, \Gamma>1/\theta_{\rm j} 
\nonumber\\
L(t\ge t_a) &\sim& \pi \left( \theta_{\rm j} R\right)^2  
\Delta R^\prime \Gamma^2 j^\prime(t) \propto t^{-\alpha_3},
 \, \,\, \Gamma\le 1/\theta_{\rm j}
\end{eqnarray}
Assuming that $j^\prime(t)$ is 
with constant slope before and after $t_a$,
we have
%
%
$t^{-\alpha_3} \propto\Gamma^2 t^{-\alpha_2} $.
Therefore we find $\Gamma \propto t^{-\Delta\alpha/2}$, with 
$\Delta \alpha \equiv \alpha_3-\alpha_2$.
This behavior should be appropriate after the time
$t_*$ characterizing the start of the late prompt phase of emission.
Finally we have:
\begin{equation}
\Gamma \, =\, \Gamma_* \left({ t\over t_*} \right)^{-\Delta\alpha/2}
\end{equation}
Since $\langle\alpha_2\rangle=0.6\pm0.3$ and 
$\langle\alpha_3 \rangle = 1.25\pm0.25$ (Panaitescu 2006),
$\Delta\alpha/2$ is of the order of $0.33\pm 0.2$.
This means that the bulk Lorentz factor, at the beginning
of the late prompt phase ($t_*$), is of the order of\footnote{We use the notation
$Q_x=10^x Q$, in cgs units.}
\begin{equation}
\Gamma_*\, =\, \theta_j^{-1} \left({ t_a\over t_*} \right)^{\Delta\alpha/2}\,
\sim \, {46 \over \theta_{j,-1}}\, \left({ t_{a,4}\over t_{*.2}} \right)^{1/3}\,
\end{equation}
which is smaller than what is usually assumed for the early prompt
emission.
The barion loading of the late shells can be estimated assuming 
a given efficiency $\eta$ for the dissipation process leading to 
the radiation produced in the plateau phase.
After $t_a$, when we see the entire emitting surface,
the jet kinetic power per unit solid angle 
$L_{\rm kin}=\Gamma \dot M c^2 \sim 2 L_X/(\theta_j^2\eta)$
and then
\begin{equation}
\dot M \, =\, { 2 L(t>t_a) \over \eta \theta_j^2 \Gamma c^2}\, \propto \, 
t^{-(\alpha_2+\alpha_3)/2}
\end{equation}
which approximately gives $\dot M\propto t^{-1}$, with a large dispersion
(the dependence is the same also for the plateau phase).
The total mass in the late ejecta is relatively small, again at most
comparable with what can be estimated for the ejecta of the early prompt.
This is because, although the $\Gamma$--factors are smaller,
the total energetics of the late prompt shells is smaller than the one
of the early prompt.
This agrees with the findings of W07 that the
total radiated energy of the late shells (which is called X--ray afterglow
in that paper) is on average a
factor $\sim$10 smaller than the early prompt.
Therefore, if $\eta$ of the early and late shells is similar, 
we do not expect a big effect from the possible refreshed shocks.
We also expect that the late shells reach the front forward
shock at very different times.
This is due both because the late shells are produced at later
and later times, and also because the later the shell is produced,
the smaller its bulk Lorentz factor, and the longer the time
needed to reach the shock front which is decelerating
by the interaction with the circumburst medium.
The refreshing effect is long lasting, but diluted.
For the same reasons, we expect that the reverse shock
is also long lasting, but diluted, and therefore not 
contributing much to the total flux.
To see this,
assume for illustration a circumburst density with a wind--like profile.
Calling $t_i$ the time (after the trigger) at which a late shell is
created, and $t_d$ the deceleration time, we have that at the 
observed time $t$ the
front shock and the $i$-th late shell are 
at the distances $R_s$ and $R_i$, respectively, given by:
\begin{equation}
R_s \, =\, ct\Gamma^2 \, = \, c t \Gamma_0^2 \left( {t\over t_d}\right)^{-1/2}
\, = \, c \Gamma_0^2 (t t_d)^{1/2}
\end{equation}
\begin{equation}
R_i\, =\, c (t-t_i) \Gamma_i^2
\end{equation}
where $\Gamma_0$ is the initial Lorentz factor of the early shells.
Late shells are assumed not to decelerate
until they catch up the front shock.
Equating the two above radii we have that the $i$th shell reaches the
front shock at the time $t_c$:
\begin{equation}
t_c\, \sim \, {\Gamma_0^4\over \Gamma_i^4}\, t_d; \qquad {\rm if}~~t_c \gg t_i
\end{equation}
For $t_d=100$ s, $t_c$ ranges from 1600 s ($\Gamma_0/\Gamma_i=2$) 
to $t_c=10^6$ ($\Gamma_0/\Gamma_i=10$).
Since the decrease of the bulk Lorentz factor is associated
with a decrease of the kinetic power, we have 
some effects only for the first refreshed shocks, at times
of the order of thousands of seconds, but not later.

\subsection{The real afterglow}

In the pre--Swift era, it was generally believed that the real afterglow 
should
contain at least a  comparable amount of energy (in the emitted radiation)
of the prompt,
that it should start a few tens--hundreds of seconds after the trigger,
and that the energy band containing most of the emission is the X--ray band
(since its energy spectrum $F(\nu)\propto \nu^{-1}$ indicates that the
peak in $\nu F(\nu)$ is within or close the X--ray band).

The prediction of the start time of the afterglow seems well
confirmed in two cases: GRB 060418 and GRB 060607, as shown
in Molinari et al. (2007), thanks to ground based near IR
observation by the REM telescope.
Quite remarkably, the X--ray light curve in these two cases
does not track the near IR,
confirming that, although the afterglow theory can correctly
explain what seen in the optical, we need another component 
to explain the X--ray flux. 
We also conclude that: i) the real afterglow X--ray component 
is much weaker than thought before:
ii) the X--ray band is likely not the band where most 
of the afterglow energy is, and 
iii) the total energetics of the real afterglow is much 
smaller than thought before.

A weak afterglow can result if the microphysical parameters
$\epsilon_e$, $\epsilon_B$, at least for the first afterglow
phases, are much smaller than commonly thought.
Alternatively, the fireball can have a small kinetic energy,
as a result of a very efficient prompt phase, that was able
to convert a large fraction of the fireball energy into radiation.
Furthermore, the radiation produced by the real afterglow should 
be mostly in the IR--optical, not in the X--ray band.

There is a spectral transition between the varying and generally
hard slope of the hard X--ray emission and a more stable 
and generally softer slope of the later X--ray emission.
In W07, the distributions of the spectral
index $\beta_x$ of the prompt and the plateau phase
are broad, but a slight narrowing around a value
$\beta_x\sim 1$ for the plateau is visible.
We propose that this is not due to the prompt/afterglow
transition, but it is instead associated to the transition 
between the early prompt phase, characterized by large 
$\Gamma$--factors changing erratically, and the late prompt 
phase characterized by smaller $\Gamma$ 
decreasing monotonically.

Note also that the pre--Swift observations of the X--ray 
``afterglow'' should be re--interpreted: in many cases,
what observed even days after the trigger time 
should be late prompt, not real afterglow, emission.

\subsection{Flares}

The flares occurring mainly in X--ray (Burrows et al. 2007)
but sometimes also in the optical
light curve have been associated to internal dissipation (internal shocks)
either by late shells, or by shells produced within the first early prompt
phase, but moving with a small Lorentz factor.
In our framework, the most likely possibility is that 
a flare is produced by a late shell, moving with a somewhat larger
Lorentz factor than the shells created just earlier.
Thus there will be a chain of interactions between this
(faster--than--average) shell and the slower previous ones, 
and this mechanism can be efficient in converting the kinetic energy
of the shell into radiation.
Due to the different time Doppler contraction, late flares
should also last longer than early ones.
Alternatively, the flares could flag periods of enhanced activity of
the central engine, able for some time (i.e. tens or hundreds of seconds) 
to produce shells (or a continuous flow) of higher energy.

\section{Discussion}

The scenario here proposed allows to explain 
some GRB properties which are puzzling and mysterious.

In general, the X--ray flux can receive contributions from the
steep part of the early prompt phase, the late prompt emission, and
from the decelerating early shells which are producing the
real afterglow emission.
The late prompt emission, in the optical--UV bands, may be reduced
if the main radiation process is multiple Comptonization of UV/soft 
X--ray seed photons, 
or by self--absorption if it is synchrotron radiation.
The real afterglow emission, instead, should produce synchrotron
(and self--Compton) radiation both in the X--ray and in the optical bands.
Focusing to the plateau and later phases, we may have a complex behavior:
\begin{enumerate}
\item 
the X--ray flux is dominated by late prompt emission, while the optical
is dominated by the real afterglow. In this case the light curves in the
two bands are independent, and show no simultaneous break.
In particular, the jet break time is possibly seen in the optical,
but not in the X--rays.
\item 
Both the X--ray and the optical fluxes are dominated by late prompt emission.
In this case the two light curves are similar, they may have simultaneous
breaks (at the time $t_a$) and  the
jet break time (due to real afterglow emission) can be masked.
The dominance of the late prompt emission may however end after some
time, beyond which the real afterglow can become visible.
\item
Both the X--ray and the optical fluxes are dominated by real
afterglow emission. This is the case foreseen before Swift.
The light--curves should have an achromatic jet--break, should track
one another, and they should not show the break at $t_a$.
\end{enumerate}
The first case is sketched in Fig. 1. 
While the other two cases can also occur, the case of a late prompt 
emission dominating in the optical but not in the X--rays seems contrived.


In terms of total energetics, the scenario proposed here may be the 
least demanding for explaining what observed.
In fact, in the refreshed shock scenario, the plateau
phase is flat because the shock running into the
circumburst medium is energized by the arrival of 
shells with kinetic energies which largely overtake
the energy of the first shells, which have contributed
to the early prompt emission.
Alternatively, in the increasing $\epsilon_e$, $\epsilon_B$ 
scenario, the radiation
produced during the plateau phase is a very tiny fraction
of the carried kinetic energy.
Instead, interpreting the plateau phase as late prompt
emission, we need that the extra energy created by
the central engine in the late phase is less than
(or at most comparable to) the total energetics
of the shells responsible for the early prompt emission.
If the radiative efficiency of the early prompt is
large, we can also explain the weakness of
the real afterglow emission, since the kinetic
energy of the fireball, remaining after the 
early prompt phase, may be relatively small.


Our scenario 
is not based on a detailed model on how the central engine works. 
The following ideas should then be considered as 
speculations to be studied in future work.
After the black hole formation following the core collapse 
of the progenitor star, the equatorial core material 
which failed to form the black hole in the first place can form
a very dense accreting torus, which can sustain a strong magnetic
field, which in turn extracts the rotational energy of the
black hole.  
This accretion phase could correspond to the early prompt phase of the burst.
After this phase, some fall--back material may also be accreted. 
This phase of ``late accretion'' can last for a longer time, with a density 
of the accreting matter smaller than in the early phases. 
If so, the magnetic field that this matter can sustain is weaker than
before, with a corresponding smaller power extracted from the black hole spin.
This may well correspond to production of shells of smaller 
$\Gamma$--factors.
These shells can dissipate part of their energy with the same mechanism
of the early ones. 
Occasionally, the central engine produces a faster than average shell,  
originating the late flares often observed in the Swift/XRT light curves.


Our suggestion may be not the unique 
solution of the puzzle concerning the unpredicted behavior of the
X--ray and the optical light curves. 
Indeed, Uhm \& Beloborodov (2007)
and Grenet, Daigne \& Mochkovitch (2007) recently proposed that
the X--ray flux may be dominated by the reverse shock emission
in slow shells.
These and our proposals share the common view that the X--ray flux
can be due to a component different from what produces the optical.
The difference is that in our model the late prompt
and real afterglow emissions are completely decoupled, while
in the reverse shock scenario the two emission processes are linked.
Furthermore, in the reverse shock scenario, 
one has to assume a somewhat ad hoc time profile of the $\Gamma$--factor
to explain the flat--steep X--ray transition, (Uhm and Beloborodov 2007),
which is not a simple (unbroken) power law as in our case.

Finally, there are some features that our model can predict.
Observationally, we should have the three cases
mentioned above: both the optical and the X--rays
are late prompt emission; both are real afterglow emission;
X--rays and optical are ``decoupled'', with the X--ray
due to late prompt and the optical due to real afterglow
emission, respectively.
One obvious way to check these possibilities is through
the construction of the simultaneous spectral energy 
distribution (SED), which can confirm or not 
if the X--ray and the IR--optical fluxes
belong to the same component.
The unknown extinction due to the host galaxy material
may complicate this test, but having enough photometric
data, especially in the infrared, may result in a good
determination of the extinction, and thus a 
good estimate of the extrapolation of the 
IR--optical spectrum into the X--ray range.
The SED so obtained may clearly show that the IR--optical
and X--ray emission belong (or not) to two different components.

Another test concerns the total kinetic energy
of the fireball after its radiative phase,
using the radio data, as done
e.g. for GRB 970508 by Frail, Waxman \& Kulkarni (2000).
Should the derived energetics be smaller than
what required by the refreshed shock scenario,
one could exclude this possibility, and instead
favor our scenario.

In cases in which the late prompt emission
ends, the underlying real afterglow emission 
can be revealed.
In the light curve, this should appear as a steep--flat transition
at late times (not to be confused with the usual
steep--flat--steep X--ray decay).
This can also be confirmed by the corresponding SEDs.


\acknowledgements
We thank the anonymous referee for helpful comments, 
F. Tavecchio for discussion and a 2005 PRIN--INAF grant for funding.

\clearpage

\begin{figure}
\centerline{\plotone{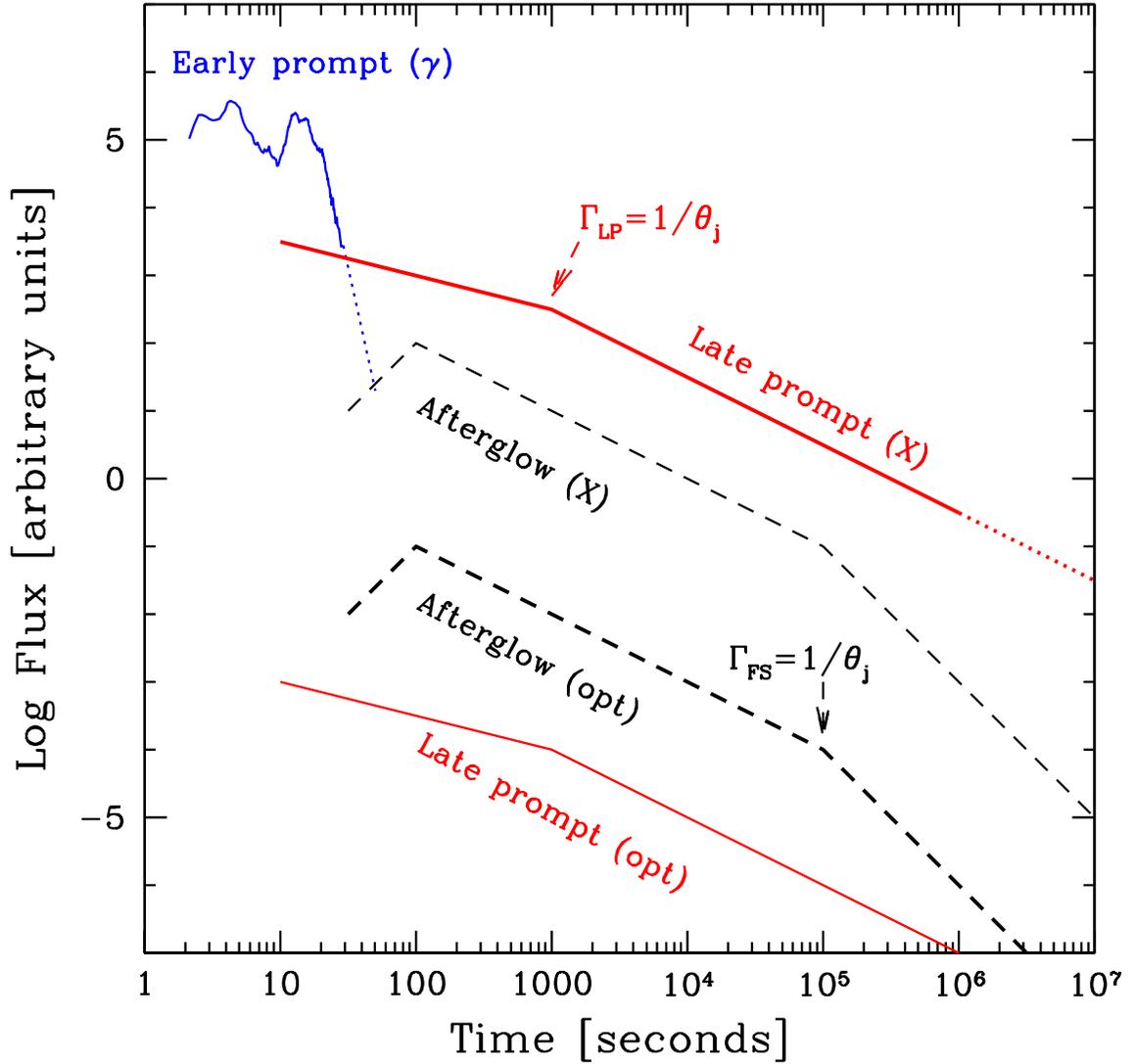}}
\caption{
Schematic illustration of the different components
contributing to the X--ray and optical light curves,
as labelled.
Scales are arbitrary. 
The case illustrated here is only one
(likely the most common) possible case (see text), when the X--ray 
flux is dominated by late prompt emission (solid line,
the dotted line corresponds to an extrapolation at very late times), 
while the optical flux is dominated by the real afterglow (dashed).
$\Gamma_{\rm LP}$ and $\Gamma_{FS}$ indicate the $\Gamma$ of the late
shells and the forward shocks, respectively.
}
\label{figura}
\end{figure}

\end{document}